\documentstyle[pre,aps,amsfonts,amssymb,eqsecnum,twocolumn,psfig]{revtex}
\begin{document}
\def\u{\bbox}
\def\mathcal#1{{\cal #1}}
\def\phi{\varphi}
\def\epsilon{\varepsilon}
\def\d{\displaystyle}

\title{An approximate renormalization-group transformation
for Hamiltonian systems with three degrees of freedom}
\author{C.\ Chandre$^1$, H.R.\ Jauslin$^1$, G.\ Benfatto$^2$, 
        and A.\ Celletti$^3$}
\address{$^1$Laboratoire de Physique-CNRS, Universit\'e de Bourgogne,
B.P.\ 47 870,\\
F-21078 Dijon, France}
\address{$^2$Dipartimento di Matematica, 
Universit\`a di Roma ``Tor Vergata'', 
Via della Ricerca Scientifica,\\ I-00133 Roma, Italy}
\address{$^3$Dipartimento di Matematica
Pura e Applicata,Universit\`a di L'Aquila, Via Vetoio,\\ 
I-67100 L'Aquila, Italy}
\maketitle

\begin{abstract}
We construct
an approximate renormalization transformation that combines 
Kolmogorov-Arnold-Moser (KAM)
and renormalization-group techniques, to analyze instabilities
in Hamiltonian systems with three degrees of freedom.
This scheme is implemented both for
isoenergetically nondegenerate and for degenerate Hamiltonians.
For the {\em spiral mean} frequency vector, we find 
numerically that the iterations of the
transformation on nondegenerate Hamiltonians tend to 
degenerate ones on the critical surface. 
As a consequence, isoenergetically degenerate 
and nondegenerate Hamiltonians belong to the same universality
class, and thus the corresponding critical invariant tori have the
same type of scaling properties. We numerically
investigate the structure of the
attracting set on the critical surface
and find that it is a {\em strange nonchaotic attractor}.
We compute exponents
that characterize its universality class.
\end{abstract}


\section{Introduction}

The breakup of invariant tori is one of the key mechanisms of the transition to 
chaos in Hamiltonian dynamics.
For two dimensional systems and for quadratic irrational frequencies, 
it has been observed that, at the transition, a
sequence of periodic orbits approaches geometrically a torus of the
given frequency, with
a nontrivial scaling behavior~\cite{greene,kadanoff,shenkerkadanoff}. 
This self-similarity has been described in terms of a nontrivial fixed point 
of a renormalization-group transformation in the case of the golden mean
\cite{mackay,mackayL,govin,chandre,cgjk,abad}.
The sequence of periodic orbits responsible for the breakup is generated 
by the continued fraction expansion of the frequency.
For the extension to 
systems with three degrees of freedom (d.o.f.) involving three incommensurate
frequencies,
we lack a theory that generalizes the continued fractions.
Numerically, three d.o.f.\ Hamiltonian systems (or equivalently four 
dimensional 
volume-preserving maps) have been studied with an extension of 
Greene's criterion
\cite{maohelleman,acs1,acs2,tompaidis1,tompaidis2} and by reconstruction of 
invariant tori using conjugation theory~\cite{bollt,kurosaki}.
The conclusion of these analyses was that there is no
geometrical accumulation of periodic orbits
around the critical torus, and thus absence of universality
(at least for the specific frequency vectors they considered).\\
In Ref.\
\cite{cj:3d}, an approximate renormalization-group scheme was described 
for a reduced family of isoenergetically degenerate Hamiltonians, which
is an intermediate case between two and three d.o.f., that
in appropriate coordinates can be interpreted as one d.o.f.\ system
driven by two periodic forces with incommensurate frequencies. This
was also the class of models considered in Ref.~\cite{acs1,acs2,tompaidis2} :
They studied an intermediate case between two-dimensional and
four-dimensional volume preserving maps. In particular, invariant
tori in these intermediate models act as barriers in phase space
(limiting the diffusion of trajectories) as for two d.o.f.\ Hamiltonian
systems.\\
The conclusion of Ref.~\cite{cj:3d}  was that one can still expect 
{\em universal}
behavior in the breakup of invariant tori: The universality is associated 
to a hyperbolic nonperiodic
attractor of the renormalization flow. The idea is that all
Hamiltonians attracted by renormalization to this set will display 
sequences of scaling factors that appear in a different order 
but with a universal statistical
distribution. The dominant unstable Lyapunov characteristic
exponent determines the
approach to criticality of the universality class.\\
By analyzing trajectories on the critical surface, we determine the
structure of the critical attractor. 
Our analysis indicates the existence of a {\em strange nonchaotic 
attractor} whose correlation dimension seems to have a value around 1,
but it is difficult to analyze it even numerically.\\
We take the definitions as formulated by Grebogi {\em et al.}~\cite{greb84}:
An attractor of a map is called strange if it is not a finite number of points,
nor a piecewise differentiable set. An attractor is chaotic if typical 
orbits on it have a positive Lyapunov exponent. 
The strange attractor we obtain is of the same general type as the ones
found in Ref.~\cite{greb84,feud95}, for quasiperiodically forced systems
(of one and two dimensions). Some
of their properties have been rigorously analyzed in 
Refs.~\cite{kell96,star97,star99}.\\
Chaotic attractors for renormalization maps have been conjectured 
and observed in statistical mechanics~\cite{mckay,eckmann} and 
dynamical systems~\cite{lanford,lanford2,rand,mack87}.
In Ref.~\cite{fs,satija1}, a strange chaotic attractor was 
found for renormalization
of circle maps, and in Ref.~\cite{satija2} similar evidence was found
for area-preserving maps, from the scaling analysis
of periodic orbits. The origin of randomness
in these latter studies is due to the randomness of 
the sequence of continued fraction 
approximants for an ensemble of the considered frequencies.
In contrast, in the present three d.o.f.\ case,
the rational approximants are a regular sequence, obtained
by iteration of a {\em single} unimodular matrix, which allows us to define a 
renormalization transformation with a {\em fixed} frequency vector.
We observe that this difference leads to a qualitatively distinct 
structure of the attractors. Although the geometry of the attractor is
singular (i.e., strange), it is not chaotic.\\

In the present article, we extend the approximate renormalization-group 
transformation
developed in Ref.\ \cite{cj:3d} to a more general family 
of Hamiltonians with three d.o.f. 
We find that the renormalization trajectories on the critical surface
converge to the
reduced family of Hamiltonians considered in Ref.\ \cite{cj:3d}.\\
Similar types of systems were first studied in Ref.~\cite{mms} with
an approximate renormalization scheme that kept less terms than needed to
detect an attractor. Instead, their transformation yields a rotation
(a center) which is structurally unstable as mentioned in Ref.~\cite{mms}.\\

We define a renormalization transformation which acts on 
Hamiltonians with three d.o.f.\ written in actions $\u{A}=(A_1,A_2,A_3)
\in{\Bbb R}^3$ and
angles $\u{\phi}=(\phi_1,\phi_2,\phi_3)\in {\Bbb T}^3$ (the 3-dimensional
torus parametrized, e.g., by $[0,2\pi]^3$)
\begin{equation}
\label{eqn:ham}
H(\u{A},\u{\phi})=H_0(\u{A})+V(\u{A},\u{\phi}),
\end{equation}
where $H_0$ is the integrable part of the Hamiltonian. We are interested in
the stability of the torus with frequency vector $\u{\omega}_0$.
We suppose that
this torus is located at $\u{A}=\u{0}$
for $H_0$, i.e., the linear part of $H_0$ is equal to
$\u{\omega}_0\cdot\u{A}$. 
Kolmogorov-Arnold-Moser (KAM) theorems were proven for Hamiltonians 
(\ref{eqn:ham}) (with suitable restrictions on the perturbation) provided
that $H_0$ contains a twist in at least one direction in the
actions~\cite{cj:kam}, i.e.,
the Hessian matrix 
$ \partial^2 H_0/\partial \u{A}^2$, with elements 
$ \partial^2 H_0/\partial A_i \partial A_j$,
is non-zero, and $\u{\omega}_0$ satisfies a Diophantine condition. 
It shows the existence 
of the torus with frequency vector 
${\u \omega}_0$ for a sufficiently small and smooth perturbation $V$. 
The invariant torus is a small deformation of the unperturbed one.
To assume nonzero Hessian matrix, we restrict the family of Hamiltonians
(\ref{eqn:ham}) to the ones with trace one :
\begin{equation}
\label{eqn:trace}
\mbox{tr}\left(\frac{\partial^2 H_0}{\partial \u{A}^2}\right)=1.
\end{equation}
The idea is to set up a transformation $\mathcal{R}$ that maps a 
Hamiltonian $H$ into a rescaled Hamiltonian $\mathcal{R}(H)$, 
such that irrelevant degrees of freedom are eliminated.
The transformation $\mathcal{R}$ 
should have roughly the following properties:
$\mathcal{R}$ has an attractive fixed set (trivial fixed set)
of integrable Hamiltonians that have a smooth
invariant torus with the frequency ${\u \omega}_0$. Every Hamiltonian in its
domain of attraction $\mathcal{D}$ has 
a smooth invariant torus with frequency vector ${\u \omega}_0$. 
The aim is to show that there
is another fixed set $\Lambda$ which lies on the boundary $\partial
\mathcal{D}$ (the
critical surface) and that is attractive for every Hamiltonian on 
$\partial \mathcal{D}$.\\
The transformation $\mathcal{R}$ is defined for a {\em fixed} frequency vector
${\u \omega}_0$ with
three incommensurate components. 
We choose
$
{\u \omega}_0=(\sigma^2,\sigma,1),
$
where $\sigma\approx 1.3247$  satisfies 
$\sigma^3=\sigma+1$ (named the {\em spiral mean}).
From some of its properties, $\sigma$ plays a similar role
as the golden mean in the two d.o.f.\ case~\cite{kimostlund}. 
The analogy comes from the fact that 
one can generate rational approximants by iterating a {\em single}
unimodular matrix $N$.
In what follows, we denote {\em resonance} an element of the sequence 
$\{ {\u \nu}_k=N^{k-1}{\u \nu}_1, k\geq 1\}$ where ${\u \nu}_1=(1,0,0)$ and
$$
N=\left(\begin{array}{ccc} 0 & 0 & 1 \\ 
                             1 & 0 & 0\\
			     0 & 1 & -1
            \end{array}\right).
$$
The word {\em resonance} refers to the fact that the small denominators 
${\u \omega}_0\cdot {\u \nu}_k$ appearing in the perturbation series or
in the KAM iteration, tend to zero geometrically as $k$ increases
($
{\u \omega}_0 \cdot {\u \nu}_k = \sigma^{3-k} \to 0 \mbox{ as } k\to \infty
$). We notice that $\u{\omega}_0$ is an eigenvector of $\tilde{N}$, where
$\tilde{N}$ denotes the transposed matrix of $N$.
The spectrum of $\tilde{N}$ is composed of one real eigenvalue $\sigma^{-1}$
(with $\u{\omega}_0$ as eigenvector), 
and two complex conjugated eigenvalues $\lambda_1\pm i\lambda_2
=\sqrt{\sigma}e^{\mp i\alpha}$
(which are of norm larger than one).
We denote by $\u{\Omega}_\pm=\u{\Omega}^{(1)}\pm i\u{\Omega}^{(2)}$, 
the eigenvectors
of $\tilde{N}$ associated with these complex eigenvalues.\\
Our hypothesis (which is also the starting point of a generalization
of Greene's criterion in Refs.\ \cite{acs1,acs2}) is that the sequence 
$\{\u{\nu}_k\}$ plays a leading role in the breakup of the invariant torus 
with frequency vector
${\u \omega}_0$. \\
In this paper, we study the extension of the ideas developed by 
Escande and Doveil~\cite{escandedoveil,escande,cjb}
to three d.o.f.\ Hamiltonian systems.
We build an approximate scheme by considering the three
main resonances ${\u \nu}_1$, ${\u \nu}_2$, and ${\u \nu}_3$. The 
renormalization focuses on the next smaller scale represented by the 
resonances ${\u \nu}_2$, ${\u \nu}_3$, together with
${\u \nu}_4=N{\u \nu}_3= {\u \nu}_1-{\u \nu}_3$. 
It includes a partial elimination
of the perturbation (the part which can be considered
nonresonant on the smaller
scale, namely the mode ${\u \nu}_1$), a shift of the resonances, a rescaling
of the actions and of the energy, and a translation in the action variables.
It is, in spirit, close to the type of transformations considered in
Refs.\ \cite{cgjk,abad}.\\ 
The approximations involved in this scheme are the two
main ones used by Escande and Doveil:\\
\indent a] A quadratic approximation in the actions (as the rescaled 
Hamiltonian $\mathcal{R}(H)$ is in general higher than quadratic 
in the actions).\\
\indent b] A three resonance approximation: we only keep the three main 
resonances at each iteration of the transformation, i.e.,
we consider the following family of even Hamiltonians
\begin{equation}
\label{hamiltonian}
H({\u A},{\u \varphi})=H_0({\u A})+\sum_{k=1}^3 h_k({\u A})\cos({\u \nu}_k
\cdot {\u \varphi}),
\end{equation}
where $h_k$ denotes the
amplitude of the mode $\u{\nu}_k$ of the perturbation. \\
Hamiltonian (\ref{eqn:ham}) is isoenergetically nondegenerate if the
following determinant of order 4 does not vanish
\begin{equation}
\label{cond:iso}
\mbox{det }\begin{array}{|cc|} \d \frac{\partial^2 H_0}{\partial \u{A}^2}
& \d \frac{\partial H_0}{\partial \u{A}} \\
\left(\d \frac{\partial H_0}{\partial \u{A}}\right)^T & 0 \end{array} \not= 0.
\end{equation}
On the contrary, if the determinant (\ref{cond:iso}) vanishes, 
$H$ is said to be isoenergetically degenerate.
In the following section, we define the renormalization
transformation for both cases.

\section{Renormalization transformation}

Our transformation is based on the following steps:\\
1] We apply a canonical transformation that eliminates the first main resonance
${\u \nu}_1$.
This is performed by a Lie transformation $\mathcal{U}_S:({\u \phi},{\u A})
\mapsto ({\u \phi}',{\u A}')$,
generated by a function $S({\u A},{\u \phi})$.
The Hamiltonian expressed in the new coordinates is given by
$$
H'=\exp(\hat{S})H \equiv H+\{S,H\}+\frac{1}{2!}\{S,\{S,H\}\}+\cdots,
$$
where $\{ \, , \, \}$ is the Poisson bracket between two scalar
functions of the actions and angles:
$$
\{f,g\}=\frac{\partial f}{\partial \u{\phi}} 
       \cdot \frac{\partial g}{\partial \u{A}} -
       \frac{\partial f}{\partial \u{A}} \cdot
       \frac{\partial g}{\partial \u{\phi}},
$$
and the operator $\hat{S}$ is defined as $\hat{S}H\equiv \{S,H\}$.
Denoting by $\varepsilon$ the size of $h_k$, 
the generating function $S$ is determined by the requirement that the order
$O(\epsilon)$ of the mode ${\u \nu}_1$ in $H'$ vanishes:
$$
\{S,H_0\}+h_1({\u A})\cos({\u \nu}_1\cdot{\u \phi})=0.
$$
This equation has the solution
$$
S({\u A},{\u \phi})=S_1(\u{A})\sin({\u \nu}_1\cdot{\u \phi}),
$$
where $$S_1(\u{A})=-\frac{h_1({\u A})}{{\u \omega}({\u A})\cdot {\u \nu}_1}, \quad
{\u \omega}({\u A})=\d \frac{\partial H_0}{\partial \u{A}}.$$
This step generates arbitrary orders in the action variables
[$\u{\omega}(\u{A})$ is linear in the actions]. In order to map
quadratic Hamiltonians into itself, we expand $H'$ to
quadratic order in the actions, and we neglect higher orders. The
justification for this approximation is that, as  the torus is located
at ${\u A}=\u{0}$ for $H_0$, one can expect that for small $\varepsilon$,
it is close to ${\u A}=\u{0}$. We notice that $h_2({\u A})$ and $h_3({\u A})$
are not changed up to order $O(\varepsilon^3)$.
Furthermore, we neglect all the Fourier modes except ${\u 0}$, ${\u \nu}_2$,
${\u \nu}_3$, and ${\u \nu}_4$, and all terms of order
greater than 3 in $\varepsilon$. This leads to the expression of $H'$:
\begin{eqnarray}
H'=&&H_0+h_2\cos({\u \nu}_2\cdot{\u \varphi})
+h_3\cos({\u \nu}_3\cdot{\u \varphi}) \nonumber \\
&+&\frac{1}{2}\langle \{S,h_1\cos(\u{\nu}_1\cdot\u{\varphi})\}\rangle + 
\{ S,h_3\cos(\u{\nu}_3\cdot\u{\varphi})\}, \label{eqn:H'}
\end{eqnarray}
where $\langle \, \rangle$ denotes the mean value defined as
$$
\langle h \rangle ({\u A}) = \int_{{\Bbb T}^3} \,
h({\u A},{\u \varphi})\, \frac{d^3 {\u \varphi}}{(2\pi)^3}.
$$
The last term of Eq.\ (\ref{eqn:H'}) contains
the Fourier mode ${\u \nu}_4={\u \nu}_1-{\u \nu}_3$ of amplitude 
\begin{equation}
\label{eqn:h4}
h_4({\u A})=\frac{1}{2}\left(S_1{\u \nu_1}\cdot \frac{\partial h_3 }{\partial
                       \u{A}}
                        +h_3 {\u \nu}_3\cdot \frac{\partial S_1}{\partial
			\u{A}}\right).
\end{equation}
We expand $h_4$ to quadratic order in the actions.\\
2] From Eq.~(\ref{eqn:H'}), the mean value term 
$\langle \{S,h_1\cos(\u{\nu}_1\cdot\u{\varphi})\}\rangle$ produces a linear
term in the actions. In order that the mean value of the linear term in
$H'$ becomes $\u{\omega}_0\cdot \u{A}$, we eliminate this term by a translation
in the actions 
${\u A}\mapsto{\u A}+{\u a}$, 
where $\u{a}$ is of order $O(\varepsilon^2)$ (so it does not produce any other
effect up to the second order in $\varepsilon$).\\
3] We shift the resonances ${\u \nu}_k\mapsto {\u \nu}_{k-1}$:
We require that the new angles satisfy $\cos(\u{\nu}_{k+1}\cdot\u{\varphi})=
\cos(\u{\nu}_k\cdot\u{\varphi}')$, for $k=1,2,3$. 
This is performed by the linear canonical 
transformation
$$
(\u{A},\u{\varphi})\mapsto (N^{-1}\u{A},\tilde{N}\u{\varphi}).
$$
We notice that $N$ is an integer matrix with determinant one. Therefore,
this transformation preserves the ${\Bbb T}^3$-structure of the angles.
This step changes 
the frequency ${\u \omega}_0$ into $\tilde{N}{\u \omega}_0=
\sigma^{-1}{\u \omega}_0$ (since ${\u \omega}_0$ is an eigenvector of 
$\tilde{N}$ by construction).\\
4] We rescale the energy (or equivalently the time)
by a factor $\sigma$, in order to keep
the frequency fixed at ${\u \omega}_0$.\\
5] We rescale the actions:
$$
H''({\u A},{\u \varphi})=
\lambda H'\left(\frac{{\u A}}{\lambda},{\u \varphi}\right),
$$
such that Condition (\ref{eqn:trace}) is satisfied for $H''$.
This normalization condition is essential to the convergence
of the transformation.\\
Similar type of approximate renormalization transformations has been defined 
in Ref.\ \cite{mms}. The main difference is that they used a
normalization condition such that the Hessian matrix $\partial^2H_0/\partial
\u{A} ^2$ is of rank 2, instead of Condition (\ref{eqn:trace}). 
Below, we make the distinction between degenerate and nondegenerate
Hamiltonians: we explicit the renormalization transformation in both cases.

\subsection{Isoenergetically degenerate Hamiltonians}
\label{sec:reduce}

The renormalization transformation described in this section was derived
in Ref.\ \cite{cj:3d}.
The integrable part $H_0$ is given by
\begin{equation}
\label{eqn:h0deg}
H_0({\u A})={\u \omega}_0\cdot{\u A}+\frac{1}{2}({\u \Omega}\cdot{\u A})^2,
\end{equation}
where $\u{\Omega}$ is a free vector of norm one: 
$\Vert {\u \Omega}\Vert=\left( |\Omega_1|^2+|\Omega_2|^2+
|\Omega_3|^2\right)^{1/2}=1$. In that case, the Hessian matrix
$\partial^2 H_0/\partial \u{A}^2$ is of rank one (proportional to the 
projection operator on the $\u{\Omega}$-direction); thus the 
isoenergetic determinant (\ref{cond:iso}) is zero.
The relevant direction (where there
is a twist) in the actions is $\u{\Omega}$. We expand $h_k(\u{A})$
in the ($\u{\Omega}\cdot\u{A}$)-variable:
\begin{equation}
\label{eqn:hkdeg}
h_k({\u A})=f_{k}+g_{k}{\u \Omega}\cdot {\u A}+
\frac{1}{2}m_{k}({\u \Omega}\cdot {\u A})^2.
\end{equation}
We rewrite the meanvalue terms in $H'$ of Eq.~(\ref{eqn:H'}) as
\begin{eqnarray*}
H_0(\u{A})&+&\frac{1}{2}\langle \{S,h_1\cos(\u{\nu}_1\cdot\u{\varphi})\}\rangle 
\\&=&H_0(\u{A})+\frac{1}{4}\u{\nu}_1\cdot\frac{\partial}{\partial \u{A}}
\left( S_1h_1\right)\\
&=&\u{\omega}_0\cdot\u{A}+
a{\u \Omega}\cdot {\u A}+\frac{1}{2}(1+\mu)({\u \Omega}\cdot {\u A})^2+\mbox{const}.
\end{eqnarray*}
The linear term $a\u{\Omega}\cdot\u{A}$ is eliminated by a translation in the 
actions $\u{A}'=\u{A}+\u{\Omega}a/(1+\mu)$.\\
The shift of the resonances (Step 3) changes the vector $\u{\Omega}$ into
$\tilde{N}\u{\Omega}$. In order to keep a unit norm, we define the image
of $\u{\Omega}$ by
\begin{equation}
\label{eqn:map}
\u{\Omega}'=\frac{\tilde{N}\u{\Omega}}{\Vert \tilde{N}\u{\Omega} \Vert}.
\end{equation}
The quadratic term of the integrable part of $H'$ becomes
$\sigma\Vert\tilde{N}{\u \Omega}\Vert^2(1+\mu)({\u \Omega}'\cdot{\u A})^2/2$.
We rescale the actions (Step 5) by a factor 
\begin{equation}
\label{eqn:lambda}
\lambda=\sigma\Vert \tilde{N}\u{\Omega}\Vert^2(1+\mu),
\end{equation}
such that $H_0$ is mapped into 
$$
H_0'(\u{A})=\u{\omega}_0\cdot\u{A}+\frac{1}{2}(\u{\Omega}'\cdot\u{A})^2,
$$
with $\u{\Omega}'$ given by Eq.\ (\ref{eqn:map}).
The transformation is thus equivalent to a mapping acting on an 
11-dimensional space (recall that $\u{\Omega}$ and $\u{\Omega}'$ have 
unit norm)
$$
(\{f_{k},g_{k},m_{k}\}_{k=1,2,3};{\u \Omega}) 
\mapsto
(\{f_{k}',g_{k}',m_{k}'\}_{k=1,2,3};{\u \Omega}'),
$$
defined by the following relations
\begin{eqnarray}
&& f_{k}'=\sigma^2\Vert\tilde{N}{\u\Omega}\Vert^2(1+\mu) f_{k+1}, 
\label{eqn:ren1} \\
&& g_{k}'=\sigma\Vert\tilde{N}{\u\Omega}\Vert g_{k+1},
 \\
&& m_{k}'=\frac{1}{1+\mu}m_{k+1} 
\qquad \mbox{ for } k=1,2 \\
&& f_{3}'=\sigma^2\Vert\tilde{N}{\u\Omega}\Vert^2(1+\mu) h_4^{(0)},  \\
&& g_{3}'=\sigma\Vert\tilde{N}{\u\Omega}\Vert h_4^{(1)},\\
&& m_{3}'=\frac{2}{1+\mu}h_4^{(2)}, \label{eqn:ren2}
\end{eqnarray}
where $h_4^{(i)}$ is the coefficient in $(\u{\Omega}\cdot\u{A})^i$ of $h_4$
given by Eq.\ (\ref{eqn:h4}). Denoting by
$$
\beta_1=\frac{\u{\Omega}\cdot\u{\nu}_1}{\u{\omega}_0\cdot\u{\nu}_1}=
\frac{\Omega_1}{\sigma^2},
$$
and
$$
\beta_3=\frac{\u{\Omega}\cdot\u{\nu}_3}{\u{\omega}_0\cdot\u{\nu}_1}=
\frac{\Omega_3}{\sigma^2},
$$
we obtain explicit expressions for $\mu$ and $h_4^{(i)}$ of the renormalization
map:
\begin{eqnarray*}
&& \mu=\frac{3}{2}\beta_1(g_1-\beta_1f_1)(\beta_1g_1-\beta_1^2f_1-m_1),\\
&& h_4^{(0)}=-\frac{1}{2}\left[ \beta_3f_3(g_1-\beta_1f_1)+\beta_1f_1g_3
\right],\\
&& h_4^{(1)}=-\frac{1}{2}\left[ ( (\beta_1+\beta_3)g_3-2\beta_1\beta_3f_3)
(g_1-\beta_1f_1)\right.\\
&&\qquad \qquad \quad \left. +\beta_1f_1m_3+\beta_3f_3m_1\right],\\
&& h_4^{(2)}=-\frac{1}{2}\left[ ( (\beta_1+\beta_3/2)m_3-\beta_1(\beta_1+2
\beta_3)g_3+3\beta_1^2\beta_3f_3)\right. \\
&& \qquad \qquad \quad \times (g_1-\beta_1f_1)\\
&& \qquad \qquad \quad \left. +(\beta_1/2+\beta_3)m_1g_3
-3\beta_1\beta_3m_1f_3/2\right].
\end{eqnarray*}
Iterating the renormalization map, Eq.~(\ref{eqn:map}) reduces to 
a rotation; in fact, as $\u{\omega}_0$ is an eigenvector of $\tilde{N}$ with an
eigenvalue of norm smaller than one, the $\u{\omega}_0$-direction of
the vector $\u{\Omega}$ is contracted. The renormalization transformation
reduces to a 10-dimensional map, where the vector $\u{\Omega}$ rotates in 
the plane $(\u{\Omega}^{(1)},\u{\Omega}^{(2)})$. It is thus parametrized
by an angle $\theta$ defined by
$$
\u{\Omega}=\rho( \u{\Omega}^{(1)}\cos \theta+\u{\Omega}^{(2)}\sin \theta ).
$$
If we choose $\u{\Omega}^{(i)}$ such that
\begin{eqnarray*}
&&\u{\Omega}^{(1)}=(\sigma^{-1/2}\cos \alpha, 1, \sigma^{1/2}\cos \alpha),\\
&&\u{\Omega}^{(2)}=(\sigma^{-1/2}\sin \alpha, 0, -\sigma^{1/2}\sin \alpha),
\end{eqnarray*}
where $\lambda_1\pm i\lambda_2=\sigma^{1/2}e^{\mp i\alpha}$ are the two complex
conjugated eigenvalues of $N$,
the expression of $\u{\Omega}$ becomes
$$
\u{\Omega}=\rho(\sigma^{-1/2}\cos(\alpha-\theta), \cos\theta, 
\sigma^{1/2}\cos(\alpha+\theta)),
$$
where, since $\Vert\u{\Omega}\Vert=1$, $\rho=[F(\theta)]^{-1/2}$, and
\begin{equation}
\label{eqn:F}
F(\theta)=\sigma^{-1}\cos^2(\alpha-\theta)+\cos^2\theta
+\sigma\cos^2(\alpha+\theta).
\end{equation}
The parameters $\beta_1$ and $\beta_3$ are expressed as functions of $\theta$:
\begin{eqnarray*}
&& \beta_1=\sigma^{-5/2}\frac{\cos(\alpha-\theta)}{[F(\theta)]^{1/2}},\\
&& \beta_3=\sigma^{-3/2}\frac{\cos(\alpha+\theta)}{[F(\theta)]^{1/2}}.
\end{eqnarray*}
The expression of the norm $\Vert\tilde{N}{\u\Omega}\Vert$ is given by
$$  
\Vert \tilde{N}\u{\Omega}\Vert = \left(\sigma\frac{F(\alpha+\theta)}{F(\theta)}
\right)^{1/2}.
$$

\subsection{General quadratic Hamiltonians}
\label{sec:quadra}

For the most general quadratic Hamiltonians, we consider the
family
\begin{equation}
\label{eqn:quadra}
H(\u{A},\u{\phi})=f(\u{\phi})+
[\u{\omega}_0+\u{g}(\u{\phi})]\cdot\u{A}+\frac{1}{2}\u{A}\cdot
[M+m(\u{\phi})]\u{A},
\end{equation}
where $M$ and $m$ are $3\times 3$ symmetric matrices, and $\u{g}$ is a vector. 
The
matrix $M$ is assumed to be nonzero
(its trace is equal to one) and $\u{g}$ is not parallel to 
$\u{\omega}_0$. In Step 3, the vector $\u{g}$ is renormalized into
$\tilde{N}\u{g}$. Thus the iterations of $\tilde{N}$ converge to the 
plane defined by $\u{\Omega}^{(1)}$ and $\u{\Omega}^{(2)}$, i.e.,
the $\u{\omega}_0$-direction of the perturbation is contracted
(as the modulus of the eigenvalue associated with $\u{\omega}_0$ 
is lower than one). 
In $H_0$, except the term $\u{\omega}_0\cdot\u{A}$ which is kept
fixed by renormalization, in all the other terms of higher order in $\u{A}$,
the $\u{\omega}_0$-direction is also contracted.
Notice that the quadratic term can be written as
$$
\frac{1}{2}\u{A}\cdot M\u{A}=\frac{1}{2} \sum_{i,j=1,2,3}
m_0^{(i,j)}(\u{\Omega}^{(i)}\cdot\u{A})(\u{\Omega}^{(j)}\cdot\u{A}),
$$
where $\u{\Omega}^{(3)}=\u{\omega}_0$, and 
$\u{\Omega}^{(1)}$ and $\u{\Omega}^{(2)}$ are the real and imaginary
part of the eigenvectors $\u{\Omega}_\pm$ of $\tilde{N}$.
Then it is sufficient to consider perturbations which
only depend on the variables $(\u{\Omega}^{(1)}\cdot\u{A})$ and 
$(\u{\Omega}^{(2)}\cdot\u{A})$, and an integrable part $H_0$ of the form
\begin{equation}
\label{eqn:h0nondeg}
H_0(\u{A})=\u{\omega}_0\cdot \u{A} +\frac{1}{2} \sum_{i,j=1,2}
m_0^{(i,j)}(\u{\Omega}^{(i)}\cdot\u{A})(\u{\Omega}^{(j)}\cdot\u{A}),
\end{equation}
where $\u{\Omega}^{(i)}$ is a fixed vector (real or imaginary part of
the complex eigenvectors of $\tilde{N}$), and the matrix $m_0$, with
elements $m_0^{(i,j)}$, is symmetric and will be a variable of the 
renormalization 
map with the restriction that the trace of the 
Hessian matrix is equal to one. There are
two directions $\u{\Omega}^{(1)}$ and 
$\u{\Omega}^{(2)}$ of twist in the actions. 
The Hessian matrix is noninvertible, but, in general,
the isoenergetic determinant (\ref{cond:iso}) is nonzero.
We write $h_k$, for $k=1,2,3$, in the
$(\u{\Omega}^{(i)}\cdot\u{A})$-variables:
\begin{eqnarray}
h_k(\u{A})&=&f_{k}+\sum_{i=1,2} g_{k}^{(i)} (\u{\Omega}^{(i)}\cdot\u{A})
\nonumber \\
&&+\frac{1}{2}\sum_{i,j=1,2}
m_{k}^{(i,j)}(\u{\Omega}^{(i)}\cdot\u{A})(\u{\Omega}^{(j)}\cdot\u{A}),
\label{eqn:hknondeg}
\end{eqnarray}
where the matrices $m_{k}$, whose elements are $m_{k}^{(i,j)}$,
are symmetric. We expand $h_4$ given by Eq.\ (\ref{eqn:h4}) such that
Eq.\ (\ref{eqn:hknondeg}) defines also the coefficients of the Taylor 
expansion of 
$h_4$.
The shift of the resonances (Step 3) changes $\u{\Omega}^{(1)}$ and 
$\u{\Omega}^{(2)}$ into
\begin{eqnarray*}
&&\tilde{N}\u{\Omega}^{(1)}=
        \lambda_1\u{\Omega}^{(1)}-\lambda_2\u{\Omega}^{(2)},\\
&&\tilde{N}\u{\Omega}^{(2)}=
        \lambda_2\u{\Omega}^{(1)}+\lambda_1\u{\Omega}^{(2)}.
\end{eqnarray*}
This is equivalent to a rotation in the plane 
$(\u{\Omega}^{(1)},\u{\Omega}^{(2)})$ combined with an amplification 
by a factor $(\lambda_1^2+\lambda_2^2)^{1/2}=\sqrt{\sigma}$.
This step changes the matrix $m_{k}$ into $m_{k}'$ whose
elements are
\begin{eqnarray*}
m'^{(1,1)}_{k}&=&\lambda_1^2 m^{(1,1)}_{k}+2\lambda_1\lambda_2
m^{(1,2)}_{k}+\lambda_2^2 m^{(2,2)}_{k},\\
m'^{(1,2)}_{k}&=&m'^{(2,1)}_{k}\\ &=&
-\lambda_1\lambda_2 m^{(1,1)}_{k}+(\lambda_1^2-\lambda_2^2)
m^{(1,2)}_{k}+\lambda_1\lambda_2 m^{(2,2)}_{k},\\
m'^{(2,2)}_{k}&=&\lambda_2^2 m^{(1,1)}_{k}-2\lambda_1\lambda_2
m^{(1,2)}_{k}+\lambda_1^2 m^{(2,2)}_{k}.
\end{eqnarray*}
The matrix $m_0$ is changed into $m'_0$ according to the same formulae.
The vector $\u{g}_{k}=(g_k^{(1)},g_k^{(2)})$ 
is mapped into $\u{g}'_{k}=(g'^{(1)}_k,g'^{(2)}_k)$ whose elements
are
\begin{eqnarray*}
&& g'^{(1)}_{k}=\lambda_1 g_{k}^{(1)}+\lambda_2 g_{k}^{(2)},\\
&& g'^{(2)}_{k}=-\lambda_2 g_{k}^{(1)}+\lambda_1 g_{k}^{(2)}.
\end{eqnarray*}
The rotation (under the action of $\tilde{N}$) of $\u{g}_{k}$ and 
$m_{k}$ is analogous to the rotation of $\u{\Omega}$ [see Eq.\
(\ref{eqn:map})]; the amplification is compensated by the rescaling 
of the actions to 
avoid divergences of the transformation. We rewrite the mean-value term
in $H'$ as
\begin{eqnarray*}
\langle \{S,h_1\cos(\u{\nu}_1\cdot\u{\varphi})\}\rangle &=&\mbox{const}+
\sum_{i=1,2}a^{(i)}{\u \Omega}^{(i)}\cdot {\u A}\\
&& +\sum_{i,j=1,2}\mu^{(i,j)}({\u \Omega}^{(i)}\cdot {\u A})
({\u \Omega}^{(j)}\cdot {\u A}).
\end{eqnarray*}
The linear terms $a^{(i)}{\u \Omega}^{(i)}\cdot {\u A}$ are eliminated by a 
translation in the actions. The mean-value of the quadratic part of
$H'$ is $\sum_{i,j=1,2}\sigma(m'^{(i,j)}_0+\mu^{(i,j)})
({\u \Omega}^{(i)}\cdot {\u A})
({\u \Omega}^{(j)}\cdot {\u A})/2$,
and thus the new Hessian matrix is given by
$\sum_{i,j=1,2}\sigma(m'^{(i,j)}_0+\mu^{(i,j)}){\u \Omega}^{(i)}\otimes
{\u \Omega}^{(j)}$, where the elements of the matrix ${\u \Omega}^{(i)}\otimes
{\u \Omega}^{(j)}$ are $({\u \Omega}^{(i)}\otimes
{\u \Omega}^{(j)})_{kl}={\Omega}^{(i)}_k {\Omega}^{(j)}_l$. 
In order to have the trace of the Hessian
matrix of the rescaled Hamiltonian equal to one, we rescale the actions (Step 5)
by a factor 
$$
\lambda=\sigma\sum_{i,j=1,2}(m'^{(i,j)}_0+\mu^{(i,j)})
\mbox{tr}(\u{\Omega}^{(i)}\otimes\u{\Omega}^{(j)}).
$$
The approximate transformation is equivalent to a mapping
acting on a 20-dimensional space (since the matrices $m$ are symmetric
and $m_0$ has a constant trace):
\begin{eqnarray*}
&& (\{f_{k},g^{(i)}_{k},m^{(i,j)}_{k},
m_0^{(i,j)}\}_{k=1,2,3;i,j=1,2})\\
&&\qquad \qquad \mapsto
(\{f''_{k},g''^{(i)}_{k},m''^{(i,j)}_{k},
m''^{(i,j)}_0\}_{k=1,2,3;i,j=1,2}),
\end{eqnarray*}
defined by the following relations
\begin{eqnarray*}
&& f''_{k}=\lambda\sigma f_{k+1}, \\
&& g''^{(i)}_{k}=\sigma g'^{(i)}_{k+1},\\
&& m''^{(i,j)}_{k}=\frac{\sigma}{\lambda} m'^{(i,j)}_{k+1},\\
&& m''^{(i,j)}_0=\frac{\sigma}{\lambda} m'^{(i,j)}_0,
\end{eqnarray*}
for $ k=1,2,3$ and $i,j=1,2$.

\section{Renormalization flow}

For each scheme (Secs.\ \ref{sec:reduce} and \ref{sec:quadra}),
the numerical implementation shows that there are
two main domains separated by a {\em critical surface}:
one where the iteration converges towards a family of integrable 
Hamiltonians (trivial fixed set), and the other
where it diverges to infinity. 

\subsection{Reduction to degenerate Hamiltonians}

As a first result, we numerically observe that the transformation 
acting on nondegenerate Hamiltonians considered in Sec.\ \ref{sec:quadra}
tends to the degenerate ones of Sec.\ \ref{sec:reduce},
{\em on the critical surface}. 
More precisely, if we consider Hamiltonians
(\ref{eqn:quadra}) with $M$ of rank 3, the contraction in the 
$\u{\omega}_0$-direction, as explained in Sec.\ \ref{sec:quadra}, reduces
the rank of $M$ by one; the $3\times 3$ matrix $M$ is thus reduced to a
$2\times 2$ matrix $m_0$.
Furthermore the numerical results show that the
renormalization reduces this rank to one
when we iterate on the critical surface.
Figure 1 shows the evolution,
under the renormalization map, of the determinant of $m_0$. 
The upper
curve corresponds to a starting Hamiltonian
in the domain of attraction of the trivial fixed set,
and the lower one corresponds to iterations on the critical surface
(both evolutions start with the same quadratic part).
We also check that the determinant of the matrices $m_{k}$ tends
to zero. Furthermore, the directions of the vectorial parameters
($\u{g}$ and $m$) tend to
be aligned by the iteration : Hamiltonians (\ref{eqn:hknondeg}) 
tend to Hamiltonians
(\ref{eqn:hkdeg}), and Hamiltonian (\ref{eqn:h0nondeg}) tends to
Hamiltonian (\ref{eqn:h0deg}), all with a same direction $\u{\Omega}$. In 
order to characterize this, we define $\u{\Omega}$ as the unit vector
with the same direction as $\u{g}_{1}$. We compute the norm
of $\u{\Omega}_2-\u{\Omega}$ and $\u{\Omega}_3-\u{\Omega}$, where
$\u{\Omega}_2$ (resp.\ $\u{\Omega}_3$) is a unit vector with direction
$\u{g}_{2}$ (resp.\ $\u{g}_{3}$). Moreover, 
in order to see that the quadratic terms are proportional to
$\u{\Omega}\otimes\u{\Omega}$, we compute the norm of 
$m_0-c\u{\Omega}\otimes\u{\Omega}$ where $c$ is defined by the constant
trace of $m_0$ (similar calculations have been done
for the other matrices $m_{k}$). These differences tend to zero
as we iterate a Hamiltonian on the critical surface (they are of
order $10^{-5}$ after 20 iterations on the critical surface).\\
From these observations, we conclude that isoenergetically degenerate 
and nondegenerate Hamiltonians belong to the same universality
class. According to the general renormalization-group picture,
the corresponding critical invariant tori are predicted to have the
same type of scaling properties.\\
We lack an explanation
of the mechanism of this second reduction of the rank. 
We remark that this second reduction is not just an effect of the rescaling
(steps 3 to 5) as it is the case for the ${\u \omega}_0$-contraction :
the second reduction does not happen outside the critical surface.
We conjecture that this reduction will also occur in an exact
renormalization scheme, but this point has not yet been explored.\\
We remark that the choice of the normalization condition (\ref{eqn:trace})
seems essential to obtain a nontrivial attractor. Other choices~\cite{mms},
like $\mbox{det }m_0=1$, do not lead to a critical attractor and
the iterations appear to diverge on the critical surface (one 
eigenvalue of $m_0$ tends to zero and the other one to infinity).\\
The reduction to rank one
allows us to work, for the precise analysis of the attractor,
with the data obtained for the degenerate case (Sec.\ \ref{sec:reduce}).

\subsection{Trivial attractor}

The domain of attraction of the trivial fixed set
is the domain where the perturbation of the iterated Hamiltonians tends to
zero. However, the renormalization trajectories in this domain do not converge 
to a fixed
Hamiltonian but converge to a {\em smooth quasiperiodic} set of integrable
Hamiltonians. 
This can be explained by looking
at the map (\ref{eqn:map}). The eigenvalues of $\tilde{N}$ are 
$\sigma^{-1}$ and
$\sqrt{\sigma}e^{\pm i\alpha}$ where 
$\alpha\approx 2\pi \times 0.3880$ ($\alpha=\arccos(-\sigma^{3/2}/2)$).
The map (\ref{eqn:map}) leads asymptotically to a rotation of angle $\alpha$
in the plane $(\u{\Omega}^{(1)},\u{\Omega}^{(2)})$,
after a contraction in the $\u{\omega}_0$-direction, as explained in
Sec.~\ref{sec:reduce}.
The values of the rescalings (\ref{eqn:lambda})
at the trivial fixed set are given by
a smooth function of $\theta$. It is given explicitely by
$$
\lambda(\theta)=\sigma^2\frac{F(\alpha+\theta)}{F(\theta)},
$$
where $F$ is given by Eq.~(\ref{eqn:F}).
This trivial rescaling
curve is depicted in Fig.~2.
Since $\alpha/2\pi$ is close to $7/18$, the evolution of $\lambda$ oscillates 
approximately with period 18.

\subsection{Critical attractor}

On the critical surface, the renormalization flow 
converges to an attracting set. This set has a codimension 1 stable manifold,
i.e., one expansive
direction transverse to the critical surface.
This set plays, for the system we consider, the same role as the nontrivial
fixed point of the renormalization-group transformation for quadratic
irrational frequencies in two d.o.f.\ Hamiltonian systems.
In particular, its existence implies 
{\em universality} for one-parameter families crossing the critical surface.
Different trajectories of the transformation display the same values of the 
rescalings with a different order but with a {\em universal} statistical
distribution.
We define exponents that characterize the universality class
associated with the spiral mean.
The mean-rescaling is defined by
$
\lambda=\lim_{n\to \infty} \left( \prod_{j=1}^{n} \lambda_j \right)^{1/n},
$
where $\lambda_j$ is the value of the rescaling after $j$ iterations on the
critical surface. We also calculate the largest Lyapunov exponent 
$\kappa$ which measures the approach to criticality as
a function of the coupling constant, of the universality class.
The result we found is
that these limits do not depend on the Hamiltonian
on the critical surface where
we start the iteration nor on the initial choice of ${\u \Omega}$.
The coefficients $\kappa$ and $\lambda$ depend only on ${\u
\omega}_0$. Numerically, we find $\kappa \approx 0.6427$ and 
$\lambda \approx 3.1479$.\\
We provide numerical 
evidence that this attractor is strange and nonchaotic. 
We remark that Eqs.\ 
(\ref{eqn:ren1})-(\ref{eqn:ren2}) of the renormalization map have the
form of a nonlinear system $\{f_{k},g_{k},m_{k}\}$ (for $k=1,2,3$) driven
by a quasiperiodic variable $\theta \mapsto \theta +\alpha$, given by
the evolution of the vector $\u{\Omega}$ represented by Eq.~(\ref{eqn:map}).
This type of systems has been analyzed in 
Refs.~\cite{greb84,feud95,kell96,star97,star99}, where
one of the main conclusion is the existence of strange nonchaotic attractors.
In order to analyze the nontrivial attractor from this perspective,
we show in Fig.\ 2 a two-dimensional plot of the time series $(\theta_j,
\lambda_j)$ of the scaling factor $\lambda_j$, and 
the angle $\theta_j$, where $j$ is the index of
the iteration on the attractor. This figure shows that $\lambda$ appears to be
a continuous (one to one) function of $\theta$.  
The evolution of the rescaling $\lambda_j$ displays an approximate period 18 
behavior, similar to the one observed on the trivial 
attractor.
We remark that 
for the trivial attractor $\lambda(\theta)$ is smooth, while for the critical 
attractor $\lambda(\theta)$ has a set of cusps (nondifferentiable points).
Since the driving map $\theta_{j+1}=\theta_j+\alpha$ fills the circle
densely, and the renormalization map is smooth, the function $\lambda(\theta)$
must have a dense set of cusps. 
In Fig.~3, we show the corresponding plot for the parameter $g_1$.
This picture is the renormalization trajectory of a single initial
Hamiltonian on the attractor. It shows that $g_1(\theta)$ is not a single valued
function of $\theta$. Similar pictures are obtained for the other coordinates
of the map. \\
In order to analyze the structure of the attractor in more detail, we consider
it from a different point of view: We take a set of initial conditions on the
critical surface (not on the attractor) parametrized by an angle $\theta$
varying over a small interval $[\theta_1,\theta_2]$. Figure~4a shows 
the projection of this segment on the plane $(\theta,g_1)$. Figure~4b shows 
the image
of the projection after 100 iterations, and Fig.~4c
after 350 iterations. This shows that as the segment comes closer to the attractor
the number of steep oscillations becomes larger, suggesting that in the limit they
correspond to discontinuities of the attractor. This behavior is similar to the 
observations of Ref.~\cite{greb84} and is compatible with the results of 
Ref.~\cite{star99}. These oscillations are associated to abrupt changes of 
the signs
of the coordinates. One can conjecture that on the attractor, there is an infinite
number of such changes of sign. The same kind of phenomena is observed for the other
coordinates. In order to see the effect of these changes of sign in the renormalization
dynamics on the attractor, we compare in Fig.~5 the power spectrum
of the time series of $g_1$, with the one of $|g_1|$. The second one indicates that the
behavior of $|g_1|$ is very close to quasiperiodic (with frequencies 1 and $\alpha/2\pi$),
while $g_1$ shows a broad spectrum.\\
A further insight into the structure of the attractor can be obtained by following
the trajectories for a single initial $\theta$ and a set of different choices of the
other coordinates at a given ``time'' (i.e., after a fixed number of iterations of 
the map).
After the relaxation transient, one observes that
all initial points fall into one of eight points of the attractor (in the example of 
Ref.~\cite{greb84}, there are two such limiting points). The iteration of these
eight points generates eight branches of the attractor that are not completely
disjoint. In order to visualize these eight branches, we display in Fig.~6
the values of an observable $\mathcal{B}$ that allows to distinguish them. A suitable
choice is the weighted sum of the coordinates of $\u{x}=(x_1,\ldots,x_9)$
$$
\mathcal{B}(\u{x})=\sum_{i=1}^9 \frac{x_i}{\eta_i},
$$
with 
$$
\eta_i=\lim_{N\to\infty}\frac{1}{N} \sum_{j=1}^N |x_i(j)|,
$$
where $x_i(j)$ is the $j$-th iterate of an initial condition of the coordinate
$x_i$.
These branches are related by symmetry, as described in Sec.~\ref{sec:sym}.\\
In summary, the numerical results suggest the following characterization
of the attractor: It is a set composed of eight branches that differ only in
the signs of one or more $h_k$ (the amplitude of the Fourier mode $\u{\nu}_k$).
Each of the branches has a dense set of discontinuities. It seems to be an example
in 9 dimensions of the type of attractors described in Ref.~\cite{greb84}.

In order to characterize the singularities, we compute the correlation 
dimension of the attractor according to the method developed by Grassberger and 
Procaccia~\cite{gp,er}. We did not find a clear result, but there is some evidence 
that the dimension has a value around 1, in agreement with Figs.~3 and 6, and with
the projection of the attractor on the plane $(g_1,m_1)$ depicted in Fig.~7.  

\subsection{Symmetries of the transformation}
\label{sec:sym}

In this paper, we have considered only even perturbations, i.e., such that
the Fourier modes are only cosine terms. The results can be extended to 
the more general case, including non-even modes $\u{\nu}_k$ ($k=1,2,3$),
by the following symmetry arguments~\cite{mackayp,chandre}.
These arguments allow also to understand the relation between the eight
branches of the attractor.\\
We analyze the effect of a shift of the origin of the angles 
on the renormalization
transformation. We denote this shift as
$$
\mathcal{T}_{\u{\theta}}: \, \u{\phi}\mapsto \u{\phi}+\u{\theta}.
$$
The KAM transformation (Step 1) commutes with $\mathcal{T}_{\u{\theta}}$. The
action of $\mathcal{T}_{\u{\theta}}$ on the shift of the resonances (Step 3) is
characterized by the following intertwinning relation~\cite{koch}:
\begin{equation}
\mathcal{R}\circ\mathcal{T}_{\u{\theta}}=\mathcal{T}_{\tilde{N}\u{\theta}}\circ
\mathcal{R},
\end{equation}
where $\mathcal{R}$ denotes the renormalization transformation. Applying
this relation to the critical attractor gives the relation between the 
renormalization trajectories for Hamiltonians $H(\u{A},\u{\varphi})$ 
and the ones for Hamiltonians $H(\u{A},\u{\varphi}+\u{\theta})$. With
this type of shift, any Hamiltonian containing only the three modes 
$\u{\nu}_k$, $k=1,2,3$ (with sine and cosine terms),
can be put into a cosine representation. Thus
the attractors for these models are directly linked by symmetries 
to the
attractor found in the even case.\\
The eight branches of the critical 
attractor are mapped into each other by symmetries of this type, realized
by shifts in the origin of the angles by
$\u{\theta}_k=\pi \u{\nu}_k$. This corresponds to the eight possible choices
of the signs of the three modes.

\section{conclusion}

This paper provides numerical results indicating 
that for the spiral mean frequency vector, the critical surface 
of the approximate renormalization transformation
is the codimension 1 stable manifold of a {\em strange nonchaotic attractor}.
This feature depends strongly on the characteristics of the 
eigenvalues of $N$. Moreover, the numerical results suggest that
the renormalization transformation can be reduced to an isoenergetically
degenerate family of Hamiltonians at criticality.
These remarks give new insights for the setup
of a systematic renormalization transformation, in the spirit of Refs.\
\cite{koch,cgjk,abad}.

\section*{acknowledgments}

We acknowledge
useful discussions with J.P.\ Eckmann, G.\ Gallavotti, H.\ Koch,
J.\ Laskar, and R.S.\ MacKay.
Support from EC Contract No.\ ERBCHRXCT94-0460 for the project
``Stability and Universality in Classical Mechanics'' is acknowledged.

\newpage
\begin{figure}
\centerline{
\unitlength=1cm
\begin{picture}(6.5,6.5)
\put(0,0.5){\psfig{figure=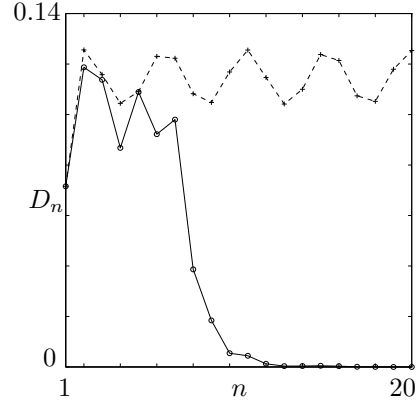,width=5.5cm,height=5cm}}
\put(0.2,0.3){1}
\put(2.5,0.3){$n$}
\put(4.6,0.3){20}
\put(0,0.7){0}
\put(-0.2,2.8){$D_n$}
\put(-0.4,5.3){0.14}
\end{picture}}
\caption{Evolution of the determinant $D_n$ of the matrix $m_0$
(as a function of the number of iterations $n$) : the
continuous line is for a trajectory of the 
renormalization transformation on the critical surface, and the
dashed line is for a trajectory inside the domain of attraction 
of the trivial fixed set.}
\end{figure}

\begin{figure}
\centerline{
\unitlength=1cm
\begin{picture}(6.5,6.5)
\put(0,0.5){\psfig{figure=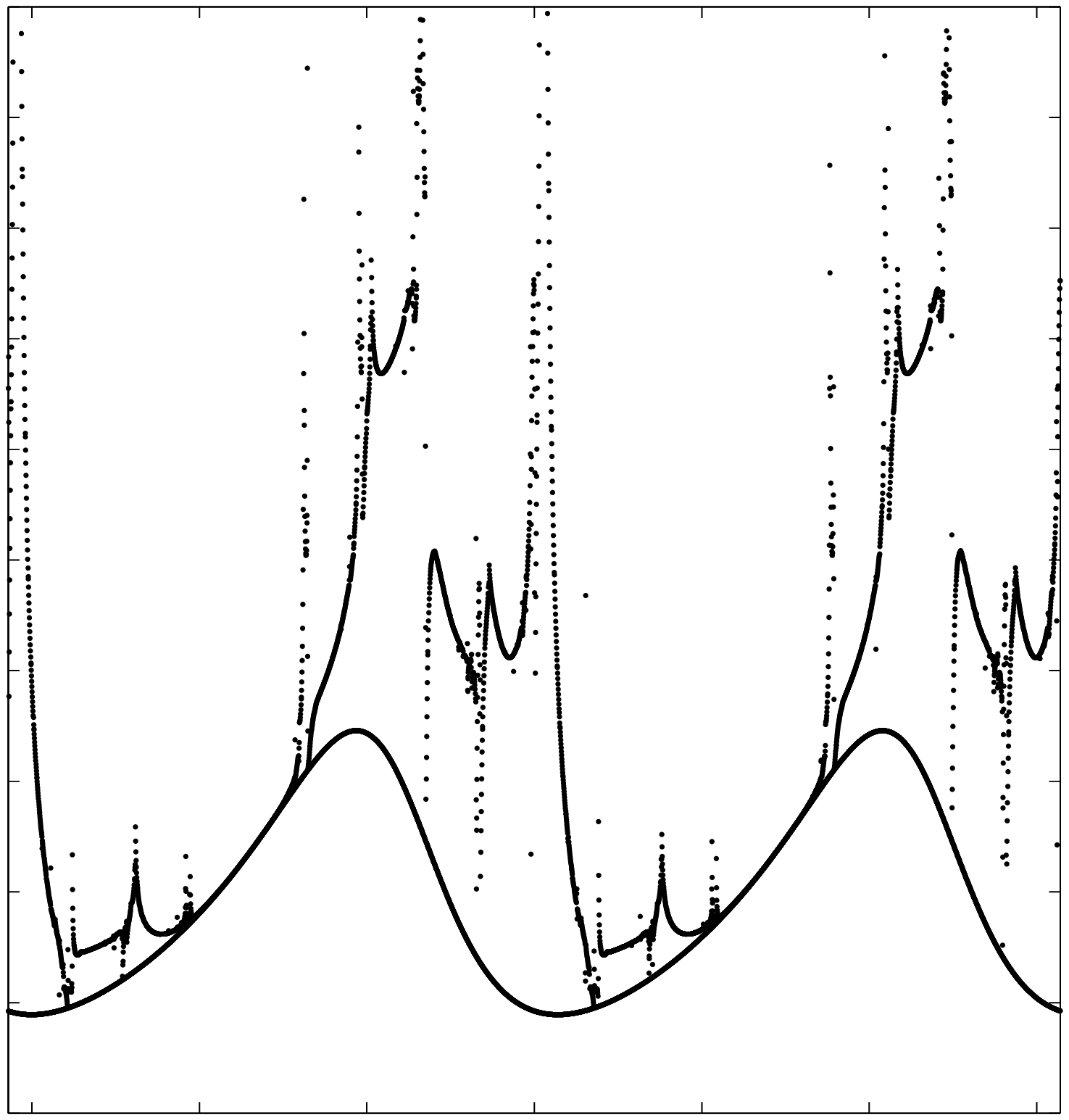,width=5cm,height=5cm}}
\put(0,0.3){$-\pi$}
\put(2.5,0.3){$\theta$}
\put(4.8,0.3){$\pi$}
\put(-0.2,0.6){0}
\put(-0.5,2.8){$\lambda(\theta)$}
\put(-0.3,4.9){10}
\end{picture}}
\caption{Values of the rescalings $\lambda$ as a function of the angle 
$\theta$
between $\u{\Omega}$ and $\u{\Omega}^{(1)}$ in the 
$(\u{\Omega}^{(1)},\u{\Omega}^{(2)})$-plane. The regular curve is for the 
trivial fixed set, and the singular curve is for the nontrivial fixed set.}
\end{figure}

\begin{figure}
\centerline{
\unitlength=1cm
\begin{picture}(6.5,6.5)
\put(0,0.5){\psfig{figure=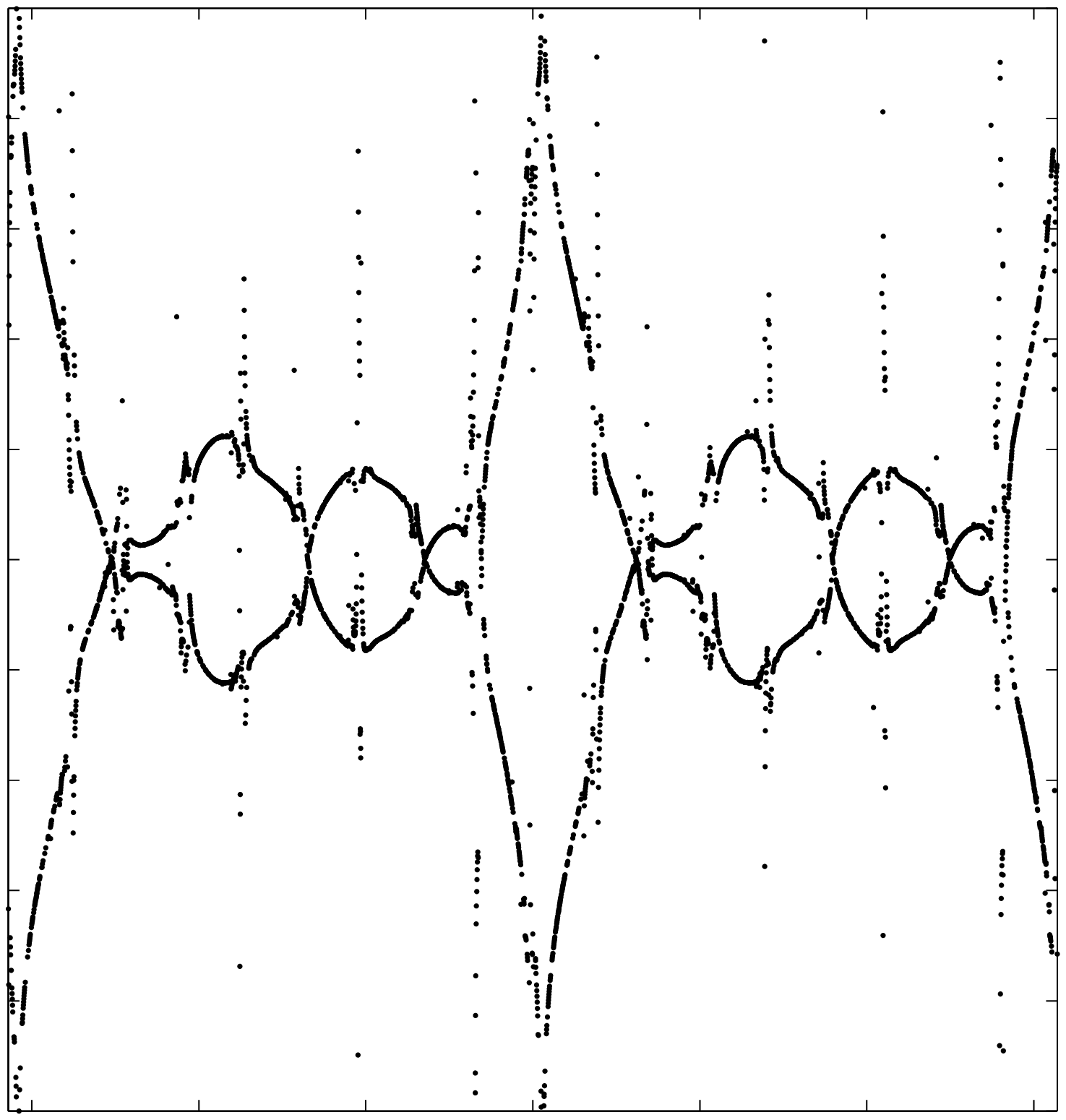,width=5cm,height=5.5cm}}
\put(0,0.7){$-\pi$}
\put(2.5,0.7){$\theta$}
\put(4.8,0.7){$\pi$}
\put(-0.2,1){-1}
\put(-0.2,3.2){$g_1$}
\put(-0.1,5.4){1}
\end{picture}}
\caption{Values of $g_1$ as a function of the angle $\theta$, on the critical 
attractor of the renormalization map.}
\end{figure}

\begin{figure}
\centerline{
\unitlength=1cm
\begin{picture}(6.5,6.5)
\put(0,0.5){\psfig{figure=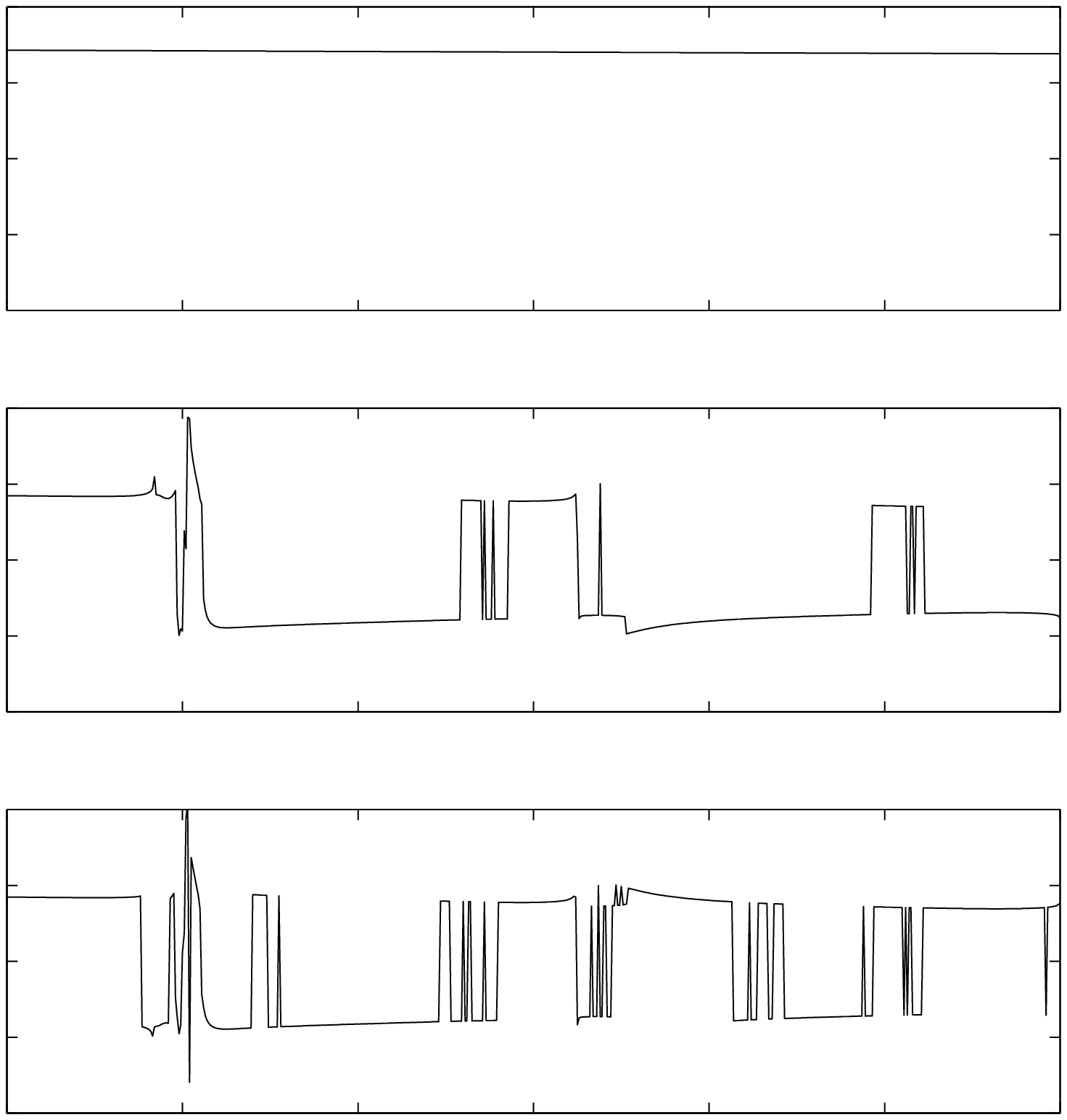,width=5.7cm,height=5cm}}
\put(0.2,0.2){6.02}
\put(2.5,0.2){$\theta$}
\put(5,0.2){6.08}
\put(0,0.6){-1}
\put(-0.2,1.2){$g_1$}
\put(0.1,1.9){1}
\put(0,2.4){-1}
\put(-0.2,3){$g_1$}
\put(0.1,3.6){1}
\put(0,4.1){-1}
\put(-0.2,4.7){$g_1$}
\put(0.1,5.3){1}
\end{picture}}
\caption{Evolution of a set of initial conditions on the critical surface,
parametrized by an angle $\theta$ varying over an interval $\left[ \theta_1,\theta_2
\right] $. We depict the projection of this set on the plane
$(\theta,g_1)$ : $(a)$ initial set,
$(b)$ after 100 iterations, and $(c)$ after 350 iterations.}
\end{figure}

\begin{figure}
\centerline{
\unitlength=1cm
\begin{picture}(6.5,6.5)
\put(0,0.5){\psfig{figure=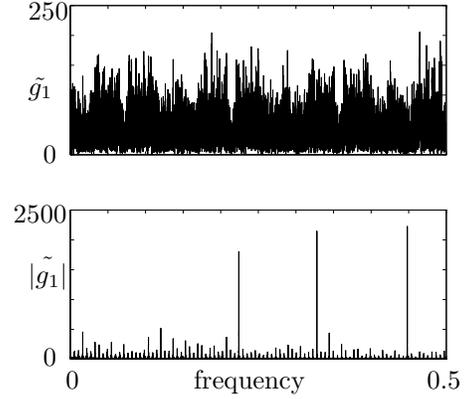,width=5.5cm,height=5cm}}
\put(0.3,0.3){0}
\put(2,0.3){frequency}
\put(5.1,0.3){0.5}
\put(0,0.6){0}
\put(-0.2,1.7){$\tilde{|g_1|}$}
\put(-0.4,2.5){2500}
\put(-0,3.3){0}
\put(-0.2,4.2){$\tilde{g_1}$}
\put(-0.2,5.2){250}
\end{picture}}
\caption{Power spectrum of the evolution on the nontrivial fixed set,
$(a)$ of the coordinate $g_{1}$, and $(b)$ of the absolute values 
$|g_1|$.}
\end{figure}

\begin{figure}
\centerline{
\unitlength=1cm
\begin{picture}(6.5,6.5)
\put(0,0.5){\psfig{figure=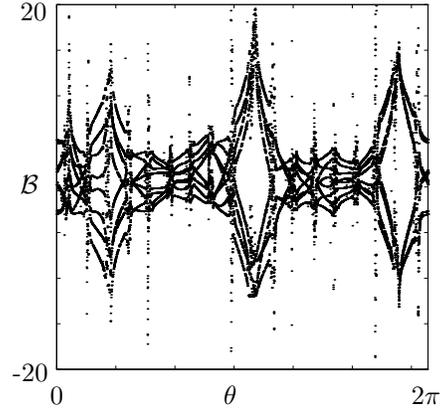,width=5.3cm,height=6cm}}
\put(0.2,0.7){0}
\put(2.5,0.7){$\theta$}
\put(5,0.7){$2\pi$}
\put(-0.3,1){-20}
\put(-0.2,3.5){$\mathcal{B}$}
\put(-0.2,5.8){20}
\end{picture}}
\caption{Values of an observable $\mathcal{B}$ as a function of 
$\theta$, on the critical attractor: it shows the eight branches of
the attractor.}
\end{figure}

\newpage
\begin{figure}
\centerline{
\unitlength=1cm
\begin{picture}(6.5,6.5)
\put(0,0.5){\psfig{figure=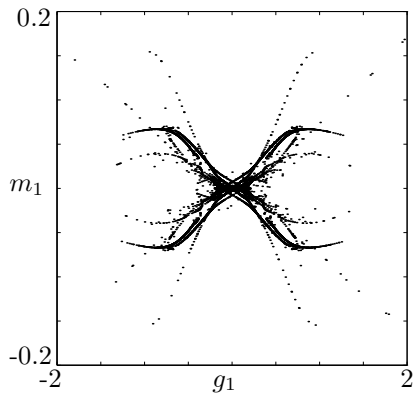,width=5.5cm,height=5cm}}
\put(0.2,0.4){-2}
\put(2.5,0.4){$g_1$}
\put(5,0.4){$2$}
\put(-0.2,0.7){-0.2}
\put(-0.2,3){$m_1$}
\put(-0.1,5.2){0.2}
\end{picture}}
\caption{Projection on the plane $(g_{1},m_{1})$
of the critical attractor of the renormalization map.}
\end{figure}


\end{document}